\documentclass[preprint,aps]{revtex4}
\newcommand{\be}{\begin{equation}}
\newcommand{\ee}{\end{equation}}
\newcommand{\ba}{\begin{eqnarray}}
\newcommand{\ea}{\end{eqnarray}}
\newcommand{\p}{\partial}
\newcommand{\f}{\frac}

\usepackage{graphicx}

\begin{document}
\title
{Quantal phase of
extreme nonstatic light waves: Step-phase evolution and its effects
  \vspace{0.3cm}}
\author
{Jeong Ryeol Choi\footnote{E-mail: choiardor@hanmail.net} \vspace{0.3cm}}
\address
{School of Electronic Engineering, Kyonggi University, 
Yeongtong-gu, Suwon,
Gyeonggi-do 16227, Republic of Korea \vspace{0.2cm}}

\begin{abstract}
\indent
The phases are the main factor that affects the outcome of various optical phenomena, such as
quantum superposition, wave interference, and light–matter interaction.
As a light wave becomes nonstatic, an additional phase, the so-called geometric phase, takes
place in its evolution. Then, due to this phase, the overall phase of the
quantum wave function varies in a nonlinear way with time.
Interestingly,
the phase exhibits a step-like evolution if the measure of nonstaticity is extremely high.
Such an abnormal phase variation
is analyzed in detail for better understanding of wave nonstaticity in this work.
As the wave becomes highly nonstatic, the phase factor of the electromagnetic wave
evolves in a rectangular manner.
However, the shape of the electromagnetic field is still a sinusoidal form
on account of the compensational variation of the wave amplitude.
The electromagnetic field in this case very much resembles that of a standing wave.
The effects accompanying the step-phase evolution, such as modification
of the probability distribution and alteration of the wave-interference profile,
are analyzed and their implications are illustrated.
\\
\\
\end{abstract}

\maketitle
\newpage

{\ \ \ } \\
{\bf 1. INTRODUCTION \vspace{0.2cm}} \\
If the parameters of a medium vary in time, the waves which propagate through it become
nonstatic according to the Maxwell's electromagnetic theory.
However, in the previous works \cite{nwh,gow,ncs,gch}, we have shown that
nonstatic quantum waves can also take place even when the medium
is transparent and its parameters do not vary.
Such nonstatic waves exhibit a peculiar time behavior, which is that
the waves expand and shrink in turn periodically in the quadrature space.

While the above interpretation for nonstatic-wave evolution is
based on the exact solutions of the Schr\"{o}dinger wave equation,
we still do not know much about the properties of the nonstatic waves
arisen in a static environment.
Many physical characteristics of the nonstatic light may long for clarification yet.
Especially, the behavior of highly nonstatic waves may greatly deviate from that
of the ordinary ones.

In this work, we are interested in the phases accompanied
the quantum wave functions of nonstatic light.
In order to interpret quantum mechanics based on particle-wave duality and non-locality,
it is necessary to know the fundamental relation between
quantum wave phase and the associated physical reality \cite{kop}.
For decades, considerable research has been devoted to phase sensitive experiments
with the waves that undergo intraatomic/intramolecular propagation \cite{spe1,spe2,spe4,spe5},
as well as free wave-packets.
The development of techniques for controlling and manipulating quantum phases
may open a new route for a variety of their applications, for example,
in wireless communications \cite{wc1,wc2}, tunable optical devices \cite{tod},
beam forming and scanning \cite{bfs-0,bfs}, radar detection \cite{rad},
and biological monitoring \cite{bmo1,bmo2}.

We know that the phase of an ordinary quantum wave in free space evolves monotonically in time.
However, if a wave becomes nonstatic, the behavior of the phase
is not so simple because there appears an additional phase (geometric phase) which shows the geometry
in the evolution of the wave \cite{gow,gch}.
The wave phase not only directly affects the electromagnetic wave phenomena,
but is responsible for a distinguishable change of the interference picture in
an electromagnetic interaction \cite{bfs,rem}.
In particular, the nature of the electromagnetic fields
in the near-field, including their phase properties, is
very different from that of the far-field in free space \cite{nef,nrp}.
Moreover, a perturbation of near-fields leads to a modification of overall phases in addition to
the change of the polarization and wave intensity \cite{cpe}.
Superposition of quantum states is also governed by the phases of its constituent substates.
The change in the probability distributions for superposition states in Hilbert space,
caused for example by the nonstaticity-induced geometric phase, is noteworthy in connection with
diverse phase-related quantum technologies \cite{gpp3,gpp4,gpp5,gpp7}.

Phases properties of the nonstatic quantum waves are analyzed in this work,
focusing especially on highly nonstatic waves.
We show that the phase evolves in a novel step-like way in time
in the limit of an extreme nonstaticity.
The associated consequences
in the wave phenomena, such as electromagnetic wave propagation, superposition of quantum
states, and wave interference, are investigated rigorously.
Physical meanings of the emergence of the abnormal step
phase are interpreted from a fundamental quantum point of view.
\\
\\
{\bf 2. OUTLINE FOR STEP-LIKE PHASE EVOLUTION \vspace{0.2cm}} \\
In this section, we see how nonstaticity of a light wave affects the properties of its phase.
The phase of a wave plays a crucial role in the description of
wave propagation, interference patterns, and quantum superpositions.
It is expected that the phase properties of nonstatic waves are
extraordinary due to the appearance of the geometric phase.
In particular, the physical characteristics of highly nonstatic waves and their quantal phases
may be significantly different from the usually known ones.

The actual electromagnetic fields are described in a complicated way via the
functions of both space and time even in free space \cite{cme2,cme3}.
Wave nonstaticity is also a factor that makes the wave representation being complicated.
To see the effects of wave nonstaticity on phase properties, we start from the formula of
the general wave functions for plane waves in a source free region.
The wave functions for such a light wave in the Fock states are given by
\be
\langle q|\psi_n \rangle = \langle q|\phi_n \rangle \exp [i\gamma_n(t) ], \label{1}
\ee
where $\langle q|\phi_n \rangle$ are the eigenfunctions and $\gamma_n(t)$ are the phases of the wave.
For a nonstatic wave, the phases are expressed as \cite{nwh,gow}
\be
\gamma_n(t) = - (n+1/2) \omega T(t) + \gamma_n (t_0),
 \label{2}
\ee
\be
T(t) = \int_{t_0}^t f^{-1}(t')dt', \label{3} \\
\ee
where $f(t)$ is a time function of the form
\be
f(t) = c_1 \sin^2 \tilde{\varphi}(t)+ c_2 \cos^2 \tilde{\varphi}(t) +
c_3 \sin [2\tilde{\varphi}(t)], \label{4} \ee
while $\tilde{\varphi}(t)=\omega (t-t_0) +\varphi$, $t_0$ and $\varphi$ are real constants.
Additionally, the coefficients in Eq. (\ref{4}) obey the conditions
$
c_1c_2-c_3^2 = 1 \label{13}
$
and $c_1c_2 \geq 1$.
For $n=0$, the phase given in Eq. (\ref{2}) is the same as that in the coherent state \cite{gch}.

\begin{figure}
\centering
\includegraphics[keepaspectratio=true]{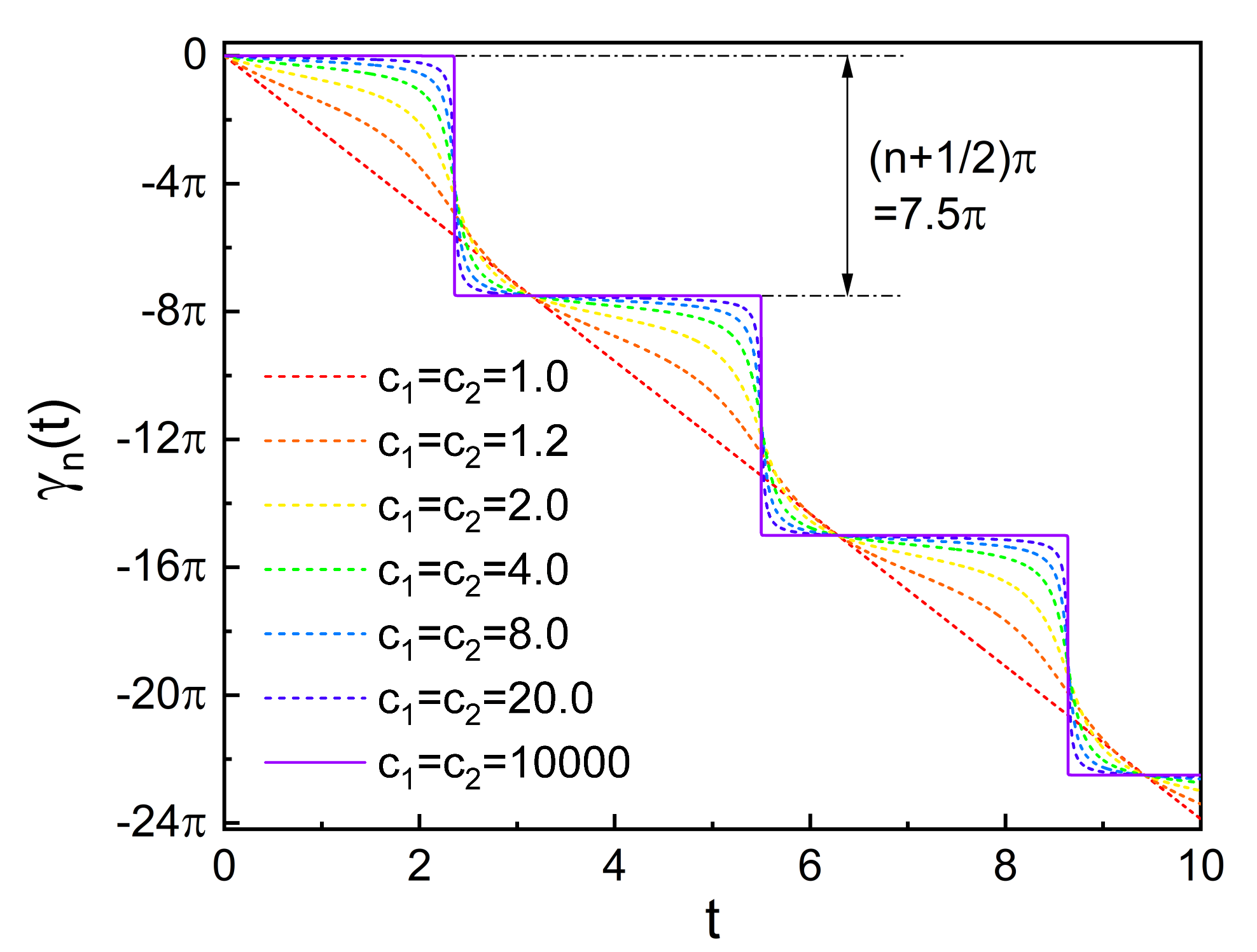}
\caption{\label{Fig1}
Evolution of the quantum phase depending on the measure of nonstaticity.
While the values of $c_1$ and $c_2$ are shown in the panel, $c_3$ is determined from the condition
$c_1c_2-c_3^2 = 1$. From this condition, two possible values of $c_3$ are allowed for a graph (one is
positive and the other is negative).
We have chosen a positive value among them for convenience: this convention will also be used
for subsequent figures.
The nonstaticity measures, $D_{\rm F}$, from red to violet curve are
0.00, 0.47, 1.22, 2.74, 5.61, 14.12, and 7071.07 in turn.
This graphic shows that the pattern of the curve for the phase evolution gradually approaches towards the step-like one
as the degree of nonstaticity increases.
We have used $\omega=1$, $n=7$, $t_0 = 0$, and $\varphi = 0$.
}
\end{figure}

Let us restrict the range of $\varphi$ in a cycle without loss of generality, such that
$-\pi/2 \leq \varphi < \pi/2$.
Then, from the integration given in Eq. (\ref{3}), we have \cite{ncs}
\be
T(t) = \Theta(t)/\omega , \label{5} \\
\ee
for $t\geq t_0$, where $\Theta(t) = \tan^{-1} Z(t) - \tan^{-1} Z(t_0)+ {\mathcal G}(t)$,
$Z(\tau)= c_3 +c_1 \tan\tilde{\varphi}(\tau)$,
${\mathcal G}(t)=\pi \sum_{m=0}^{\infty}u[t-t_0-(2m+1)\pi/(2\omega)+\varphi/\omega]$
and $u[x]$ is the unit step function (Heaviside step function).
Although our research in this work is focused on phase properties of the wave functions, the eigenfunctions
$\langle q|\phi_n \rangle$ in Eq. (\ref{1}) have also been represented in Appendix A for completeness.

\begin{figure}
\centering
\includegraphics[keepaspectratio=true]{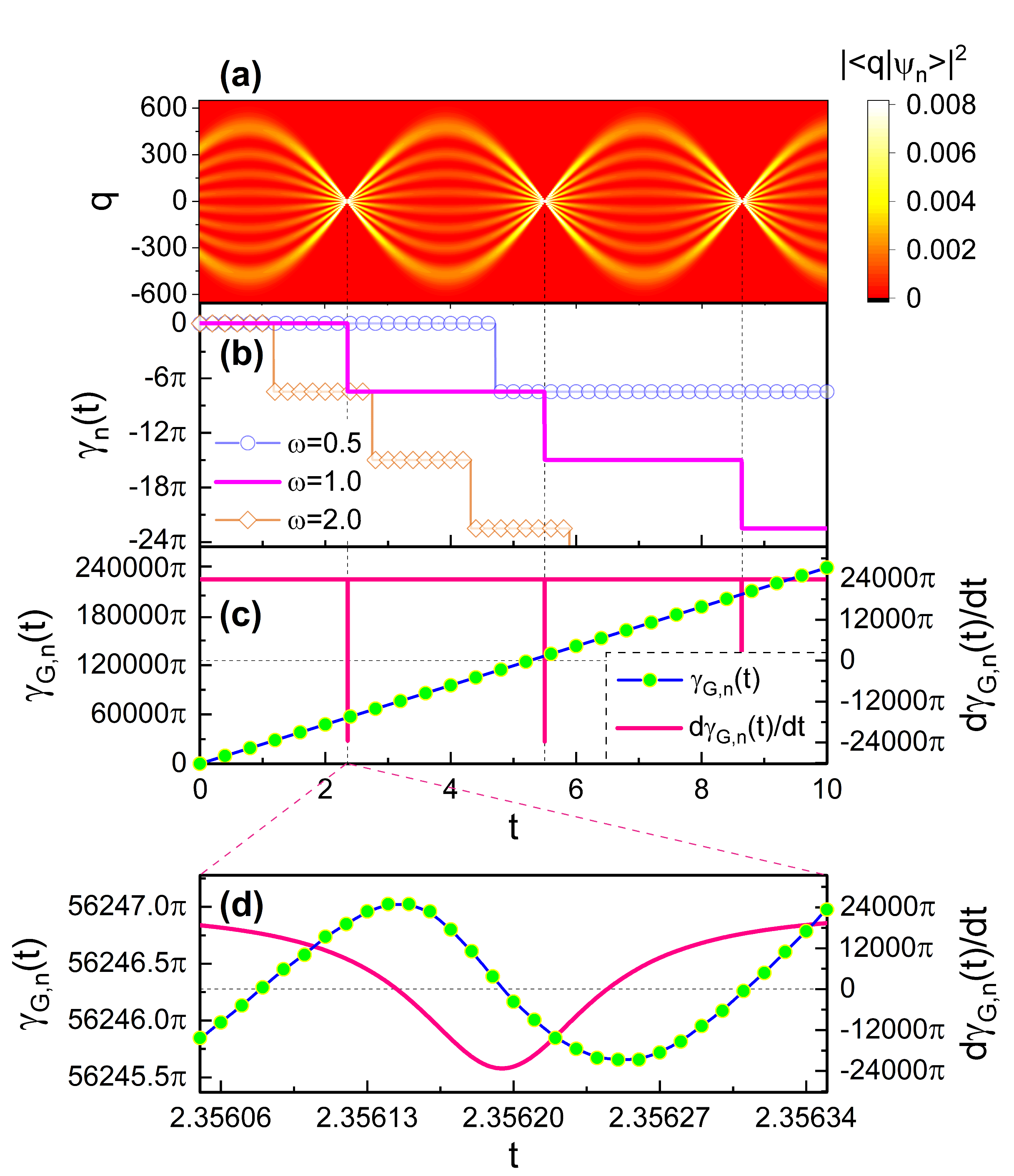}
\caption{\label{Fig2}
(a) is the density plot of the probability density $|\langle q|\psi_n \rangle|^2$ for an extreme nonstatic wave with $\omega=1$.
(b) is the step-like evolution of the quantum phase depending on $\omega$.
(c) is the evolution of the geometric
phase, $\gamma_{G,n}(t)$, and its time derivative , $d \gamma_{G,n}(t)/dt$,
for the extreme nonstatic wave with $\omega=1$.
(d) is an enlarged graphic of (c) for the time interval designated.
For the case of (c) and (d), the axis of $d \gamma_{G,n}(t)/dt$ is given in the right of the graphic.
We have used $c_1=c_2=10000$, $n=7$, $\hbar=1$, $\epsilon=1$, $t_0 = 0$, and $\varphi = 0$.
At a glance, it seems that the geometric phase in (c) varies linearly in time.
However, its gradient abruptly down whenever the quantum phase precipitates.
}
\end{figure}

Basically, we are interested in the phase of a highly nonstatic wave.
The degree of wave nonstaticity can be estimated from the measure of nonstaticity $D_{\rm F}$
that has been represented in Appendix B.
The scale of nonstaticity of a wave increases as $D_{\rm F}$ grows.
The wave nonstaticity is actually governed by the values of $c_1$ and $c_2$,
because $D_{\rm F}$ is represented in terms of them as can be seen from Eq. (\ref{22}) in Appendix B.

Equation (\ref{2}) is the same as the addition of the dynamical and geometric phases shown
in Appendix C.
The behavior of this phase when the wave nonstaticity is non-negligible may be notable
due to its characteristic abnormal properties.
Figure 1 exhibits the change of the pattern for the time evolution of the phase
as the degree of nonstaticity increases.
Except for the case $c_1=c_2=1$ which corresponds to the static wave,
the phase evolution is nonlinear.
The scale of such a nonlinearity is gradually augmented as the measure of nonstaticity increases.
Eventually, when the nonstaticity measure is extremely large, the quantum
phase exhibits a step-like time behavior, which drops periodically.
This is closely related to the time variation of the eigenfunctions given
in Eq. (\ref{qpn}) in Appendix A and the appearing
of additional phases (the geometric phases) as the wave becomes nonstatic.
The change of phase in each drop in a Fock state is $(n+1/2)\pi$ as shown in Fig. 1.

Figure 2 shows that the quantum phase precipitates whenever the probability density
constitutes a node during its time evolution in the quadrature space.
At a glance, it seems that the geometric phase depicted in Fig. 2(c) increases linearly over time.
However, it is not the case \cite{gow}.
We can see from Fig. 2(d) that the geometric phase undergoes a nontrivial nonlinear variation during
a negligibly short time.
This consequence can also be confirmed from the periodical abrupt change of the gradient of the
geometric phase, which is shown in Figs. 2(c) and 2(d) with the red curve.
This extremal behavior of the geometric phase is the extension of the usual consequence for
the nonstaticity-induced geometric phase (shown in Ref. \cite{gow}) to an extremely higher nonstatic case.
\\
\\
{\bf 3. DESCRIPTION OF NONSTATIC ELECTROMAGNETIC FIELDS \vspace{0.2cm}} \\
The analysis of spatiotemporal evolution of nonstatic light of which phase exhibits nonlinear properties
is necessary for the understanding of the propagation of such an unusual light wave.
This analysis also helps the conceptualization and characterization of
correlated coherent states for nonstatic waves, including interpreting of
the related quantum measurements.

Let us see the effects of the characteristic
phase evolution for a highly nonstatic coherent wave on its propagation.
The coherent state for a nonstatic wave is obtained by means of a generalized annihilation operator.
From the definition of such an annihilation operator in the nonstatic regime, which is
\be
\hat{a}= \sqrt{\f{\epsilon\omega}{2\hbar f(t)}}\left(1- i \f{\dot{f}(t)}{2\omega}\right) \hat{q} + i\sqrt{\f{
f(t)}{2\epsilon\omega\hbar}} \hat{p},  \label{9}
\ee
we have \cite{ncs,gch}
\ba
\hat{a} (t) = \hat{a}_0 e^{-i[\omega T(t)+\theta]} ,   \label{10}
\ea
where $\theta$ is a constant.
In particular, for $t=t_0$, this reduces to $\hat{a} (t_0) = \hat{a}_0 e^{-i\theta}$.
From the eigenvalue equation
$
\hat{a} |\alpha \rangle =\alpha |\alpha \rangle, \label{11}
$
the eigenvalue is obtained in the form \cite{ncs}
\be
\alpha(t) = \alpha_0 e^{-i[\omega T(t)+\theta]}, \label{12}
\ee
where $\alpha_0$ is an amplitude which is real.
This preliminary description for the annihilation operator and its eigenvalue
will be used later in order to unfold the theory of nonstatic coherent waves.

To see the effects of wave nonstaticity on the evolution of electromagnetic waves,
we regard the associated vector potential.
Since the vector potential is a sum of its components of each radiation mode \cite{whl}, it reads
\be
{\bf A} ({\bf r}, t) = \sum_l  {\bf u}_{l} ({\bf r}) q_{l} (t) ,  \label{13}
\ee
where $\sum_l$ is a shorthand notation for $\sum_l =\sum_{l_x=-\infty}^\infty\sum_{l_y=-\infty}^\infty
\sum_{l_z=-\infty}^\infty$ with $l_x , l_y , l_z = 0, \pm 1 , \pm 2 , \cdots$,
${\rm{\bf u}}_{l} ({\rm{\bf r}})$ is the position function and $q_{l} (t)$ is the amplitude for mode $l$.
We consider a plane wave which propagates with a periodic boundary condition,
${\rm{\bf u}}_{l} ({\rm{\bf r}})={\rm{\bf u}}_{l} ({\rm{\bf r}}+L\hat{{\rm{\bf x}}})
={\rm{\bf u}}_{l} ({\rm{\bf r}}+L\hat{{\rm{\bf y}}})={\rm{\bf u}}_{l} ({\rm{\bf r}}+L\hat{{\rm{\bf z}}})$.
Then, the position and time functions are represented, respectively, as
\be
{\bf u}_{l\nu}
({\bf r}) = \f{1}{\sqrt{V}} \hat{\bf{ \varepsilon}}_{l\nu} \exp{(\pm
i{\bf k}_l \cdot {\bf r})},  \label{14}
\ee
\be
\hat{q}_l = \sqrt{\f{\hbar f_l(t)}{2\epsilon\omega_l}}
(\hat{a}_l + \hat{a}_l^\dagger),  \label{15}
\ee
where the wave vector is of the form
${\bf k}_l = \omega_l {\bf \hat{n}}/c = (l_x\hat{{\rm{\bf x}}}
+l_y\hat{{\rm{\bf y}}}+l_z\hat{{\rm{\bf z}}})2\pi/L$,
$c$ is the speed of light in the medium, $\hat{ \bf{ \varepsilon}} _{l\nu} $ are two unit vectors
associated with the polarization direction, and $V=L^3$.
Then, the mode of the polarization is completely given from the set ($l_x$, $l_y$, $l_z$, $\nu$).

By considering only a particular mode frequency with a polarization, let us drop
the subscripts $l$ and $\nu$ from now on for convenience.
Then, the vector potential can be written such that
\ba
{\bf A} ({\bf r}, t) &=& \sqrt{\f{\hbar f(t)}{2\epsilon
V \omega}}
\big[\hat{a} e^{i{\bf k} \cdot {\bf r} }
+ \hat{a}^\dagger e^{-i{\bf k} \cdot {\bf r} } \big]. \label{18}
\ea
For the purpose of further simplicity, we now consider the case that
the wave propagates $x$ direction. Then the vector potential is reduced to 
\ba
A (x, t) &=& {\mathcal A}(t)
\cos [k x - \Theta(t)-\theta], \label{19}
\ea
where ${\mathcal A}(t) = \sqrt{{2\hbar f(t)}/{(\epsilon V \omega)}} \alpha_{0}$.
In the derivation of Eq. (\ref{19}), we used the formula given in Eq. (\ref{12}).

\begin{figure}
\centering
\includegraphics[keepaspectratio=true]{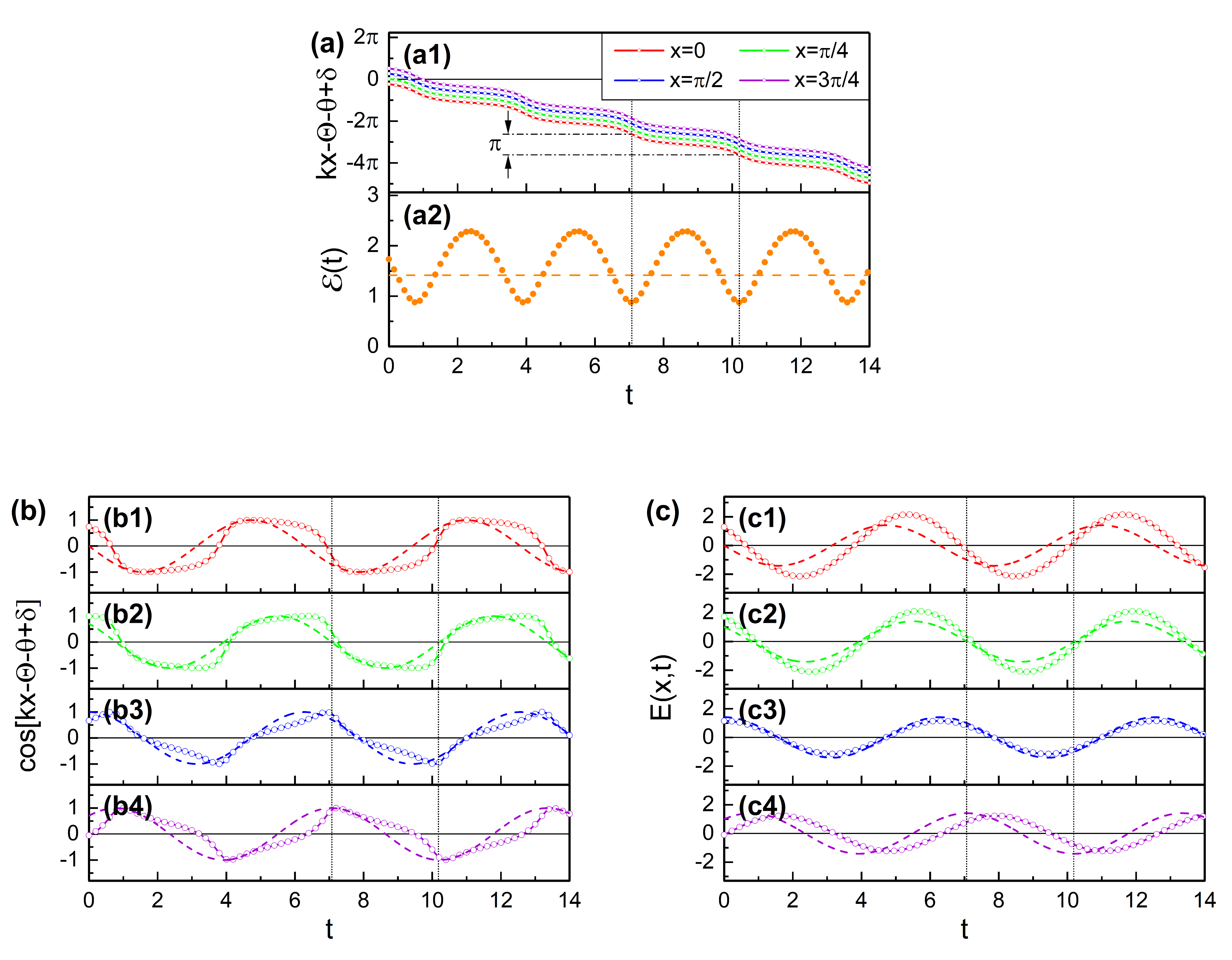}
\caption{\label{Fig3}
Graphical analyses of wave nonstaticity:
temporal evolutions of physical quantities in
conjunction with E field given in Eq. (\ref{20}), where $c_1 = c_2 =1.5$ is chosen.
Panel (a) is the phase $k x - \Theta(t)-\theta +\delta$ (a1) of E field
at several positions ($x$) designated in the legend
and the amplitude ${\mathcal E}(t)$ (orange circles in (a2));
Other two main panels are the phase factor $\cos [k x - \Theta(t)-\theta +\delta]$ (b) and E field (c)
at  $x=0$ (red circles), $x=\pi/4$ (green circles), $x=\pi/2$ (blue circles), and $x=3\pi/4$ (violet circles).
We have used $\omega=1$, $\epsilon=1$, $\hbar=1$, $k=1$, $V=1$, $\alpha_0=1$,
$t_0 = 0$, $\varphi = 0$, and $\theta=0$.
The reference lines (dashed lines) in panels (a2), (b), and (c) are the same as the main graph
of which colour is identical, but without nonstaticity ($c_1 = c_2 =1$).
}
\end{figure}

The electric and magnetic fields in the source-free space can be obtained
solely from the expression of ${\bf A}$ via the relations
\ba
& &{\bf E} = 
- \f{\p {\bf A}}{\p t}, \label{16} \\
& &{\bf B} = \nabla \times {\bf A}. \label{17}
\ea
To evaluate these equations for the wave propagating through $x$ direction,
we insert Eq. (\ref{19}) into Eqs. (\ref{16}) and (\ref{17}).
This procedure results in
\ba
E (x, t) &=&
{\mathcal E}(t) \cos[k x - \Theta(t)-\theta +\delta], \label{20} \\
B (x, t) &=& {\mathcal B}(t) \sin [k x - \Theta(t)-\theta], \label{21}
\ea
where ${\mathcal E}(t)=\omega\sqrt{1+z^2(t)}{\mathcal A}(t)/f(t)$, ${\mathcal B}(t)=-{\mathcal A}(t)k$,
$z(t) = {{\dot{f}(t)}}/{(2\omega)}$, and $\delta = {\rm atan}(-z(t),1)$.
Here, $\mu \equiv {\rm atan}(x,y)$ is the inverse function of $\tan\mu = y/x$
defined within a cycle of the angle, $0 \leq \mu < 2\pi$.
\\
\\
{\bf 4. SPATIOTEMPORAL EVOLUTION OF NONSTATIC WAVES  \vspace{0.2cm} }\\
{\bf 4.1. Analysis of mechanism for emerging nonstaticity} \\
Let us see the mechanism of appearing nonstaticity in electric field in this subsection using the development
given in the previous section.
From Fig. 3, we can confirm the emergence of nonstaticity in the electric field as the measure of nonstaticity
deviates from zero.
Both the amplitude of E field and the phase factor deviate from the normal ones as the wave nonstaticity takes place.
Figures 3(a1) and 3(a2) show that the amplitude is smallest whenever the phase drops highly.
During the amplitude evolves from its smallest one to the next smallest one,
the phase drops $\pi$.
Of course, the phase factor varies significantly when the variation of the phase is great.
However, the resultant electric field varies normally even if it deviates from the static one.
The variation of E field has been augmented
for Figs. 3(c1) and 3(c2) as the nonstaticity arose for instance,
whereas that for Figs. 3(c3) and 3(c4) has been quenched slightly.

\begin{figure}
\centering
\includegraphics[keepaspectratio=true]{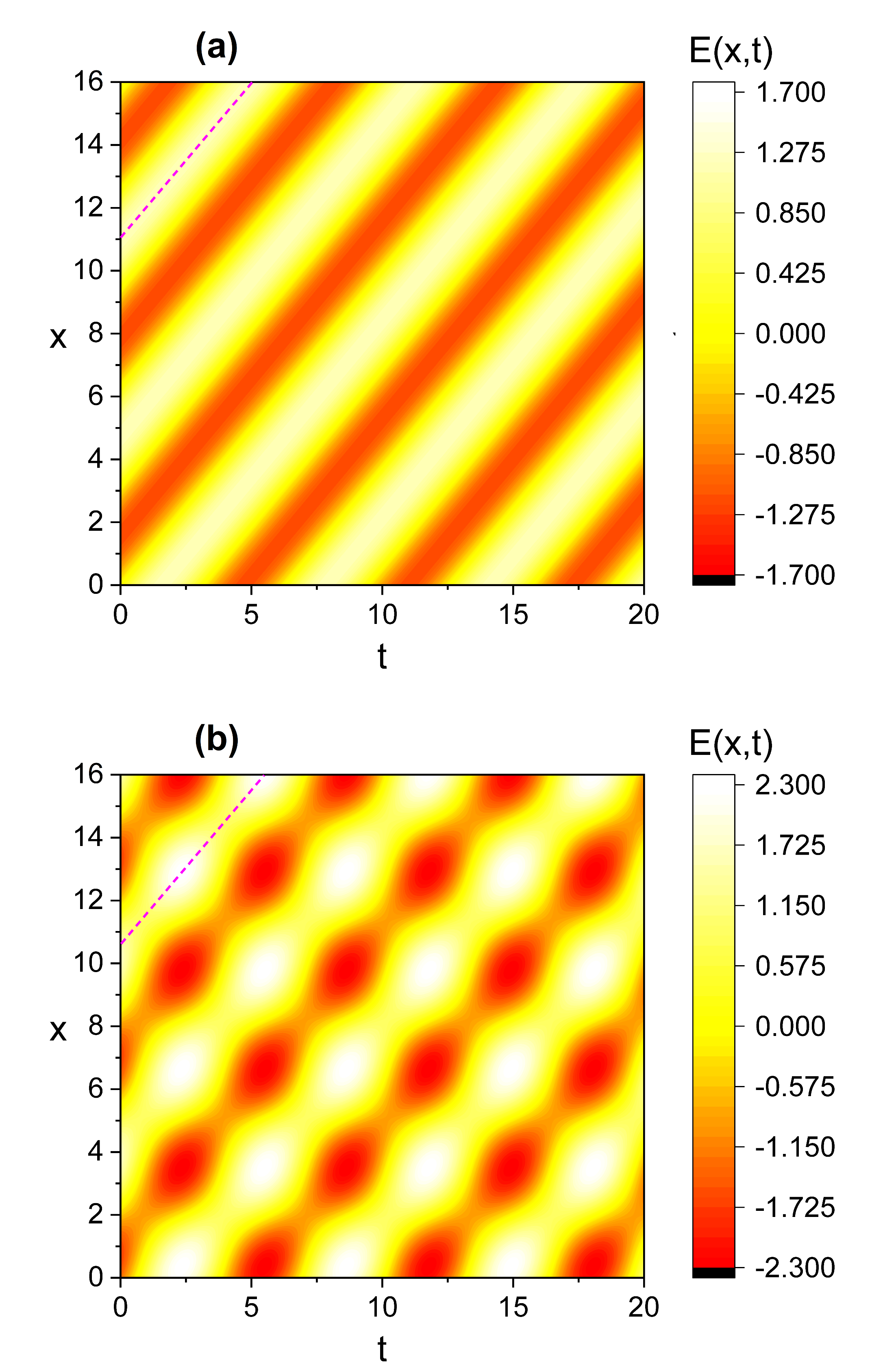}
\caption{\label{Fig4}
Density plots for the comparison of static (a) and nonstatic (b) evolutions of $E(x,t)$
where ($c_1$, $c_2$) are ($1$, $1$) for (a) and ($1.5$, $1.5$) for (b).
We have used $\omega=1$, $\epsilon=1$, $\hbar=1$, $k=1$, $V=1$, $\alpha_0=1$,
$t_0 = 0$, $\varphi = 0$, and $\theta=0$.
}
\end{figure}

\begin{figure}
\centering
\includegraphics[keepaspectratio=true]{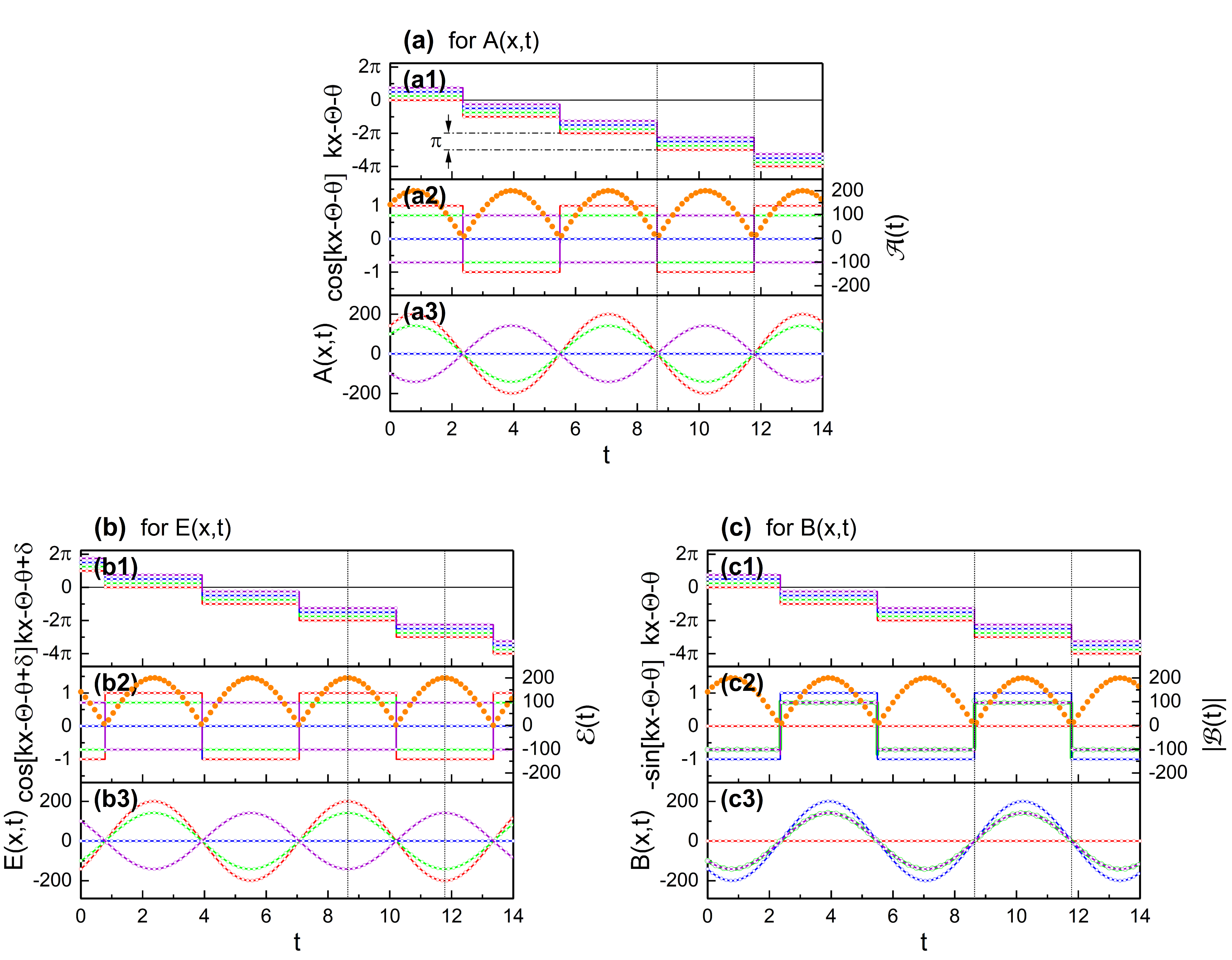}
\caption{\label{Fig5}
Graphical analyses of extreme nonstaticity from temporal evolution of related physical quantities
in conjunction with $A (x, t)$ (a), $E (x, t)$ (b), and $B (x, t)$ (c)
at $x=0$ (red lines), $x=\pi/4$ (green lines), $x=\pi/2$ (blue lines), and $x=3\pi/4$ (violet lines)
with the choice of $c_1=c_2=10000$.
Subpanels (a1), (b1), and (c1) are phases designated in Eqs. (\ref{19}), (\ref{20}),
and (\ref{21}), respectively, whereas (a2), (b2), and (c2) are phase factors in relation with them.
Extra curves composed of orange circles in (a2), (b2), and (c2) are amplitudes ${\mathcal A} (t)$,
${\mathcal E} (t)$, and $|{\mathcal B} (t)|$ that can be seen
from Eqs. (\ref{19}), (\ref{20}), and (\ref{21}), respectively.
Subpanels (a3), (b3), and (c3) are $A (x, t)$, $E (x, t)$, and $B (x, t)$ in turn.
We have used $\omega=1$, $\hbar=1$, $\epsilon=1$, $t_0 = 0$, $\varphi = 0$,
$\theta=0$, $V=1$, $\alpha_0=1$, and $k=1$.
The green line in (c2) and (c3) is enlarged so that it can be seen.
}
\end{figure}

\begin{figure}
\centering
\includegraphics[keepaspectratio=true]{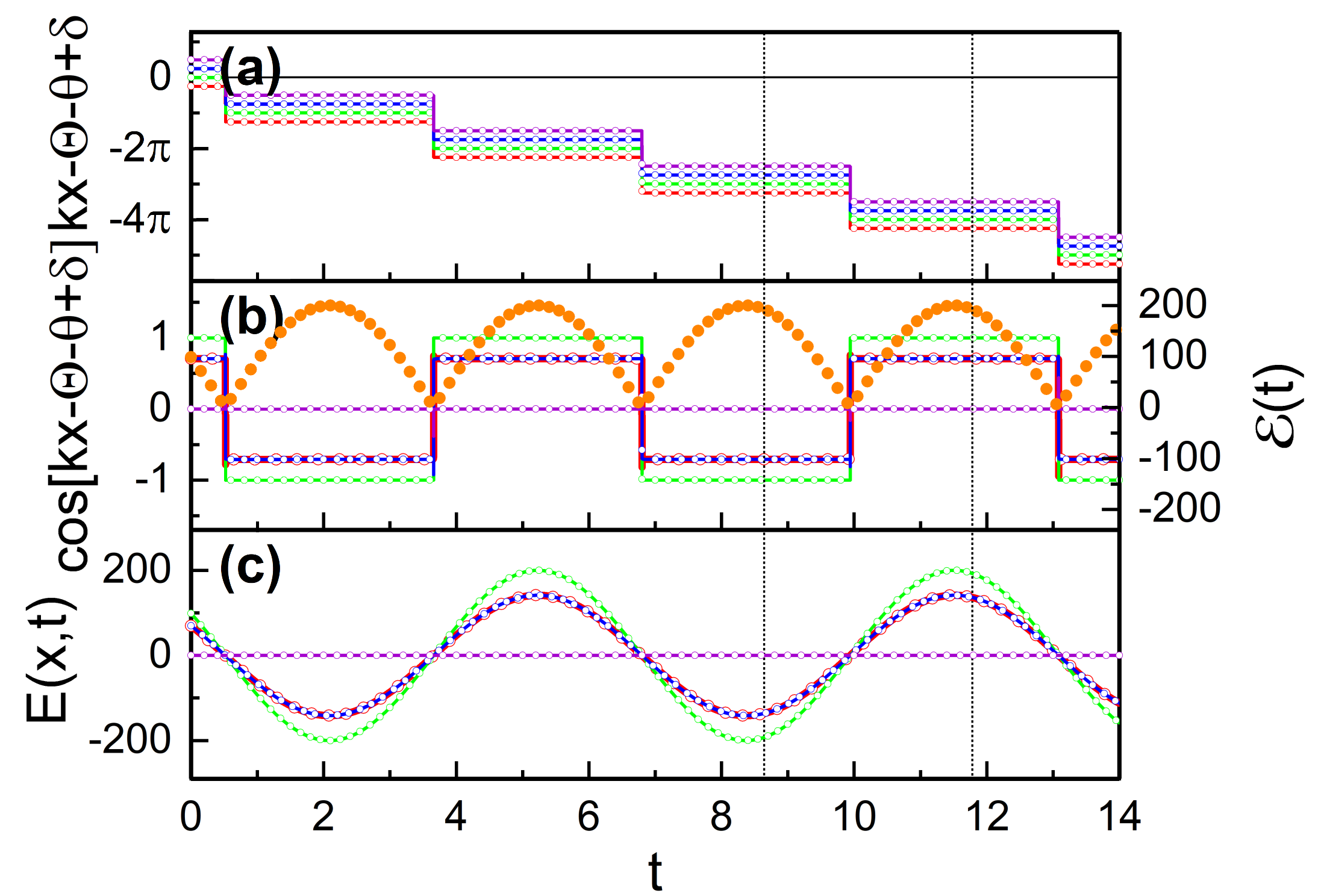}
\caption{\label{Fig6}
This is the same as Fig. 5(b) but with the choice of
($\varphi$, $\theta$)$=$($\pi/3$, $\pi/4$)
and ($c_1$, $c_2$)$=$($20000$, $1$).
}
\end{figure}

\begin{figure}
\centering
\includegraphics[keepaspectratio=true]{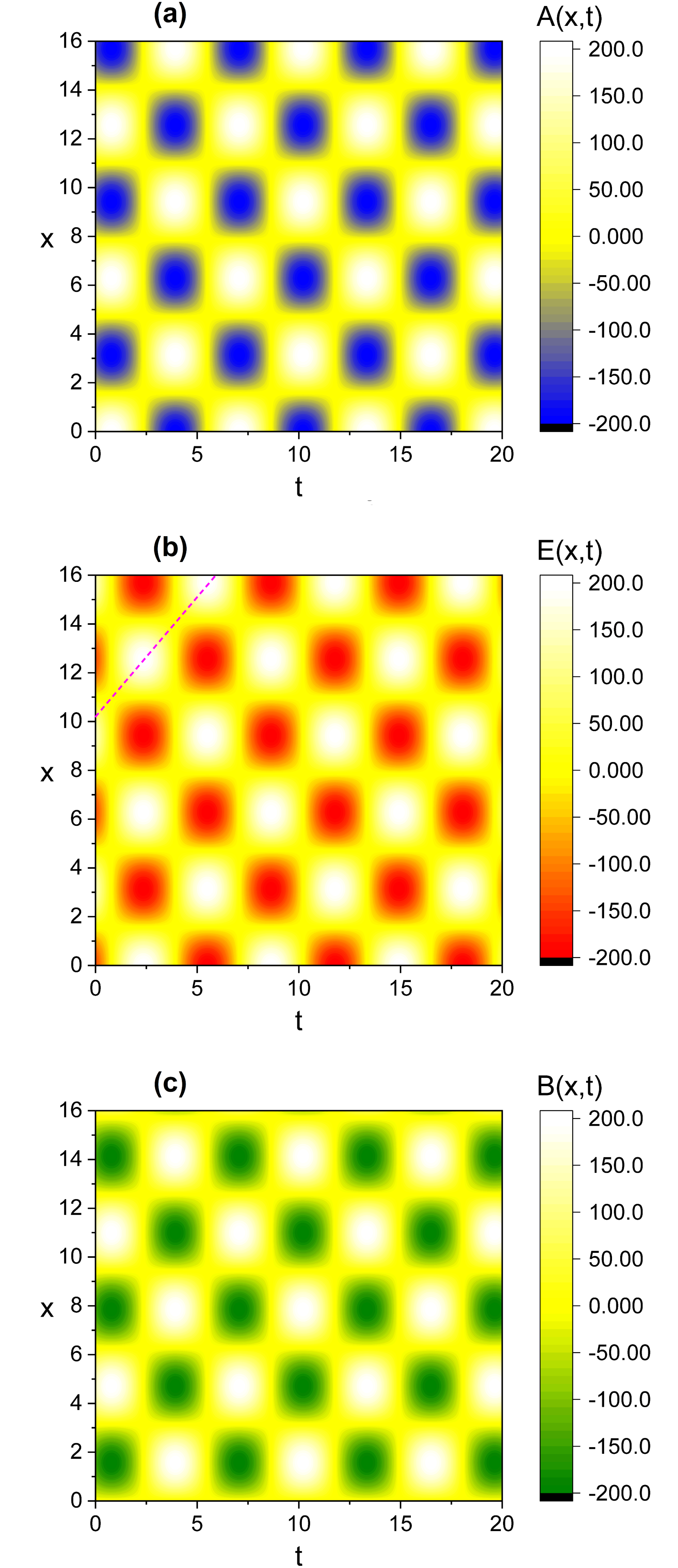}
\caption{\label{Fig7}
Density plot for the evolution of $A(x,t)$ (a), $E(x,t)$ (b), and $B(x,t)$ (c) with extreme nonstaticity,
where ($c_1$, $c_2$)$=$($10000$, $10000$) is chosen for all panels.
We have used $\omega=1$, $\epsilon=1$, $\hbar=1$, $k=1$, $V=1$, $\alpha_0=1$,
$t_0 = 0$, $\varphi = 0$, and $\theta=0$.
}
\end{figure}

Figure 4 shows the spatiotemporal evolution of the electric field, where Fig. 4(a) is static and Fig. 4(b) is nonstatic ones.
From Fig. 4(b), you can confirm that the amplitude of the field is different depending on the position $x$
as the wave becomes nonstatic.
The evolutions of the vector potential and the magnetic field for a nonstatic wave also exhibit
similar pattern as that of the nonstatic electric field.
From this discernible characteristic of nonstatic fields,
it is possible to distinguish the nonstatic electromagnetic wave from the ordinary wave experimentally.
Although we have seen only the case of electric field for brevity, the magnetic field also evolves in the same manner.
\\
\\
{\bf 4.2. Evolution of extreme nonstatic electromagnetic waves} \\
We now extend the wave evolution that we have examined in the previous subsection to
the extreme nonstatic case.
Figure 5 is the time evolution of several physical quantities related to
the vector potential and the electromagnetic
fields at equally-spaced four positions for an extreme nonstatic wave.
We see from this figure that the waves are quenched at some positions (for example, see the blue line
in Fig. 5(b3)), and active at other positions.
At a place where the waves are not completely quenched, the phase factor of a wave
evolves in a rectangular manner over time,
whereas the amplitude of the wave varies a lot.
Because the phase evolves $\pi$ during a period of amplitude variation,
the sign of phase factor changes whenever the phase drops: this is the reason for such rectangular
evolution of the phase factor.
However, the electric and magnetic fields take ordinary sinusoidal forms in their evolution.
This is due to the fact that not only the phase, but the amplitude of the wave also
varies in time. The time variation of the amplitude compensates the abnormal evolution of the
phase in a way that the resulting fields evolve sinusoidally.
If we only consider during a period of amplitude change,
it looks like that the roll of the amplitude and phase factor are changed each other.

Figure 5 together with Fig. 3 is useful for understanding the emergence of nonstaticity
in electromagnetic waves.
The variations of amplitudes shown in Fig. 3(a2) and Fig. 5(a2, b2, c2) themselves can be regarded as 
nonstaticity of waves in general.
By comparing Fig. 5(b2) with Fig. 5(c2), the amplitude of B field is smallest (highest) whenever the amplitude of E field is highest (smallest).
Such behaviors of the amplitudes are responsible for the variation of
the quadratures uncertainties in the coherent state \cite{ncs}.
In more detail, the uncertainty of $q$ quadrature is great
whenever $|{\mathcal B}(t)|$ is large,
whereas the uncertainty of $p$ quadrature is great
whenever ${\mathcal E}(t)$ is large.

We have seen the case where $\varphi = \theta$ and $c_1 = c_2$ from
the graphical analyses of the evolution of the waves until now.
However, even when $\varphi \neq \theta$ and $c_1 \neq c_2$,
the electromagnetic fields with a high nonstaticity also evolve with the same pattern given in Fig. 5.
We can confirm this fact from Fig. 6 for the case of electric field for instance.

Figure 7 is the spatiotemporal evolution of $A (x, t)$, $E (x, t)$, and $B (x, t)$
for a highly nonstatic wave. In this case, the electric and magnetic fields are very similar to
those of the standing wave.
We almost do not know the propagation direction of the light wave based on only this figure.
In what follow, we can conclude that the wave propagates along positive $x$ direction.
For the cases of Figs. 4(a,b) and 7(b), the group velocity of the wave is the gradient of the
dashed line in the figures.
\\
\\
{\bf 5. EFFECTS OF NONSTATICITY ON LIGHT-WAVE PHENOMENA \vspace{0.2cm}} \\
{\bf 5.1. Superposition states \vspace{0.2cm}} \\
Based on the principle of superposition, a light wave is allowed to be in all possible states simultaneously
until it collapses to one of the basis states by its interaction with the
environment, or by a measuring of it.
Schr\"{o}dinger's cat is a good example of such superposed states.
In fact, quantum interference between element states of a superposed state
is responsible for various nonclassical effects such as sub-Poissonian
photon statistics, oscillation of photon number distribution, and higher-order squeezing \cite{nce1,nce2}.
The quantum phase of each substate plays a major role in the formation of
the overall probability distribution that accompanied such quantum interference
in a superposition state.

\begin{figure}
\centering
\includegraphics[keepaspectratio=true]{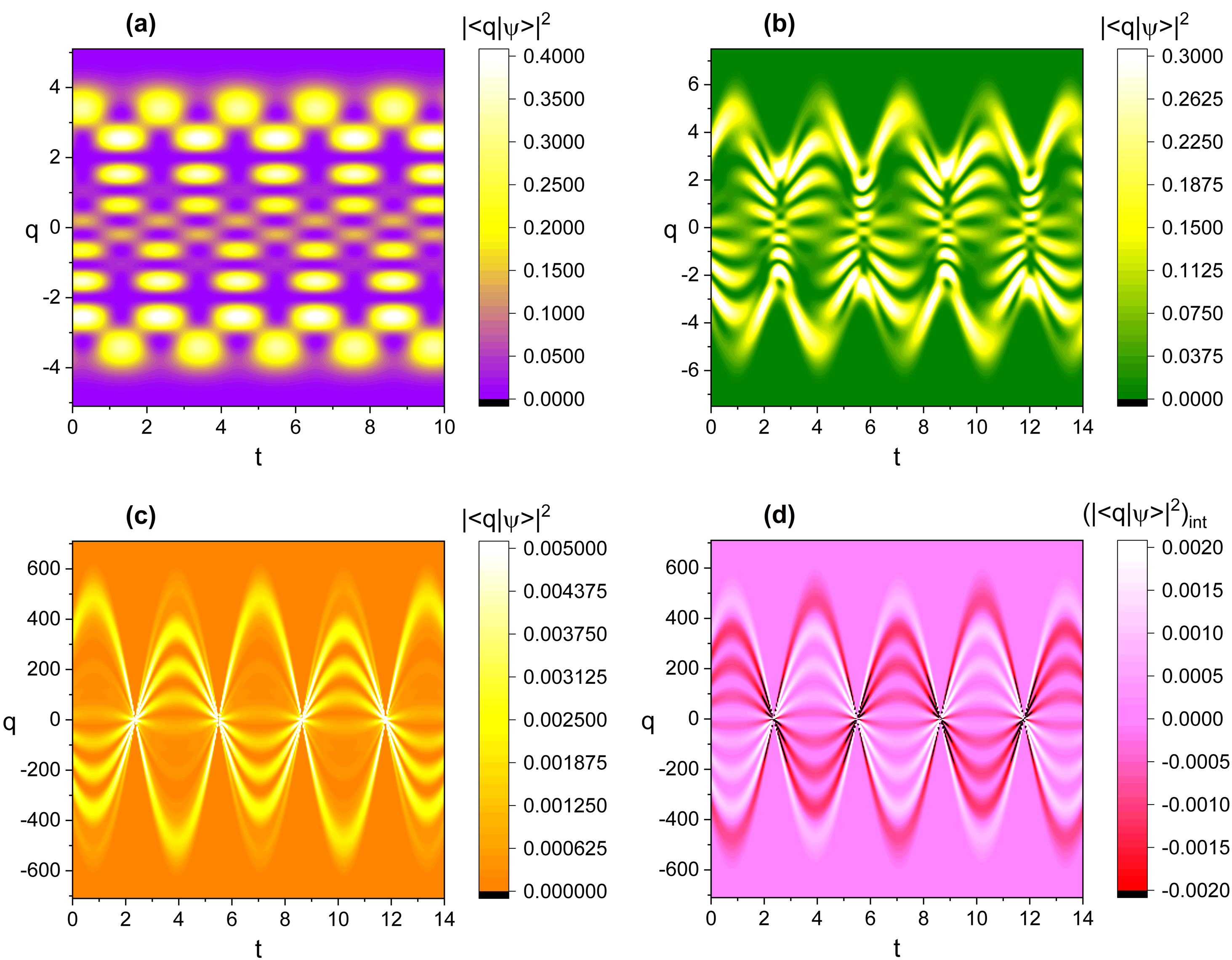}
\caption{\label{Fig8}
The evolution of the probability density for the superposition state given in Eq. (\ref{6})
with the choice of $\beta_n=1/\sqrt{2}$ and $\beta_m=(1+i)/2$.
In depicting each panel,  we used Eq. (\ref{1}) with Eqs. (\ref{2}) and (\ref{qpn}).
We have chosen ($c_1,c_2$) as (1.0,1.0) for (a), (1.5,1.0) for (b), and (10000,10000) for (c).
(d) is only for the cross term associated with (c).
While (a) is the static case, (c) is a highly nonstatic case.
We have used $\omega=1$, $n=5$, $m=8$,  $\hbar=1$, $\epsilon=1$, $t_0 = 0$, and $\varphi = 0$.
}
\end{figure}

Let us consider a superposition of two Fock states, where the resultant wave function is of the form
\be
\langle q|\psi \rangle = \beta_n \langle q|\psi_n \rangle + \beta_m \langle q|\psi_m \rangle.
 \label{6}
\ee
The complex coefficients in this representation satisfy the normalization
condition, which is $|\beta_n|^2 + |\beta_m|^2=1$.
The corresponding probability density can be written as
\be
|\langle q|\psi \rangle|^2 = |\beta_n \langle q|\psi_n \rangle|^2
+ |\beta_m \langle q|\psi_m \rangle|^2 +(|\langle q|\psi \rangle|^2)_{\rm int},
 \label{7}
\ee
where $(|\langle q|\psi \rangle|^2)_{\rm int}$ is the cross
terms which lead to interference effects and is given by
\be
(|\langle q|\psi \rangle|^2)_{\rm int} = \beta_n^* \beta_m (\langle q|\psi_n \rangle)^* \langle q|\psi_m \rangle
+\beta_n \beta_m^* \langle q|\psi_n \rangle (\langle q|\psi_m \rangle)^*. \label{8}
\ee
The phase does not affect the first and second terms in the right-hand side of Eq. (\ref{7}).
However, its effects in the third term (the cross terms) is non-negligible.
We have shown the evolution of $|\langle q|\psi \rangle|^2$ in Fig. 8.
Figure 8(c) is a highly nonstatic case whereas Fig. 8(a) is the static case.
By comparing these two panels in Fig. 8, we confirm that the probability density in the extreme nonstatic case is
quite different from that of the ordinary superposed states.
Figure 8(c) exhibits that the original superposition pattern given in Fig. 8(a) is nearly annihilated by
the effect of the characteristic nonstaticity, whereas the variation of the wave amplitude is dominant instead.
We see from Fig. 8(d) that the interference term also exhibits a similar outcome.
\\
\\
{\bf 5.2. Interference effects \vspace{0.2cm}} \\
As is well known, interference between light waves is an evidence of their wave-like property,
where the phase plays the key concept in its interpretation.
The physical effects of interference can be utilized to the development
of quantum information science and technologies related to
quantum computing, quantum metrology, and quantum communication \cite{tpi1,tpi2,tpi3,tpi4}.
When two or more nonstatic waves interact each other, the
geometric phase in each wave component may alter the interference pattern.

\begin{figure}
\centering
\includegraphics[keepaspectratio=true]{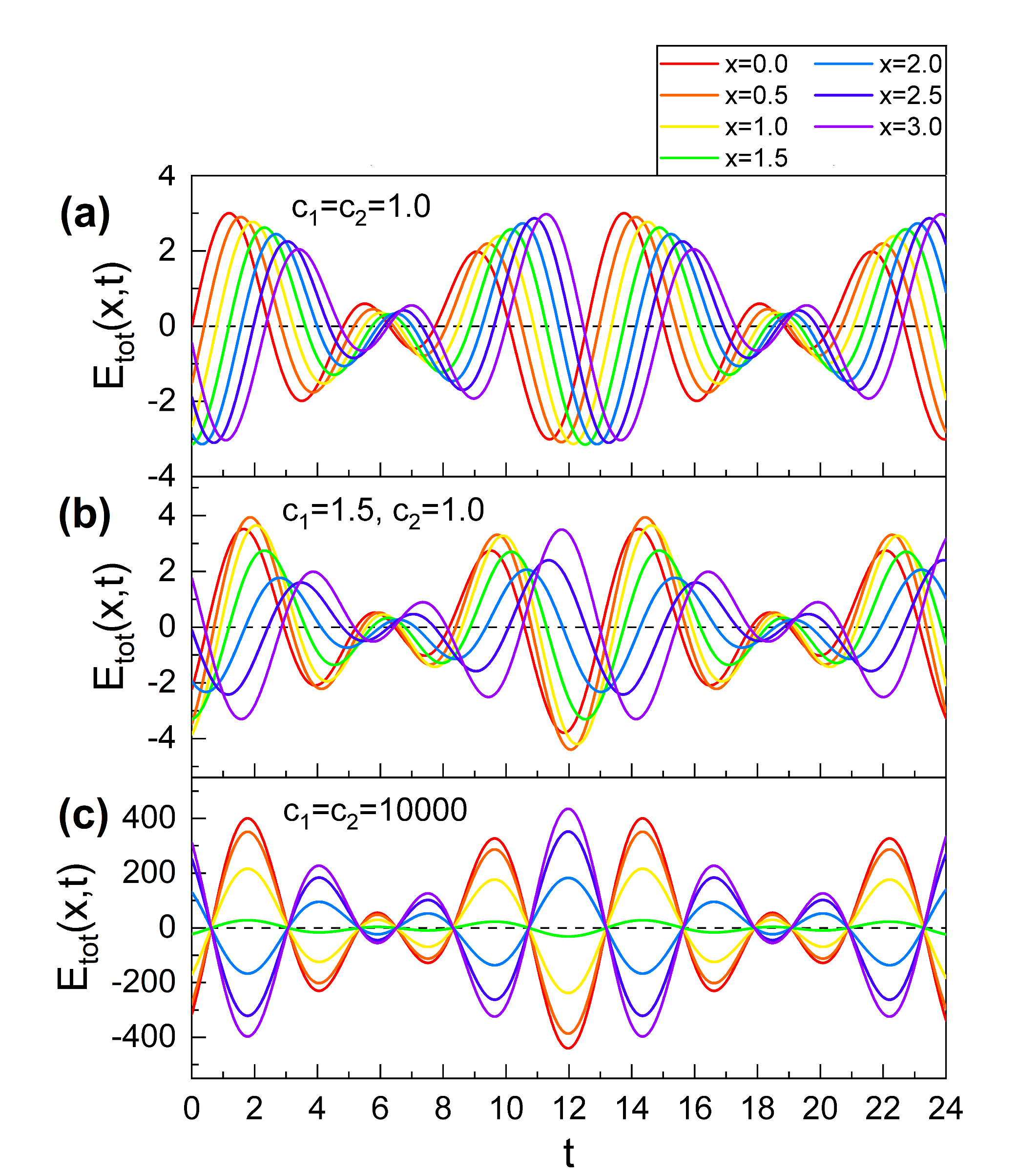}
\caption{\label{Fig9}
The time evolution of the interfered electromagnetic field given in Eq. (\ref{E215})
with the choice of $\omega_{\rm I}=1.0$ and $\omega_{\rm II}=1.5$.
All used values except for frequencies are the same for $E_{\rm I} (x, t)$ and $E_{\rm II} (x, t)$.
We considered seven different positions (see uppermost legend) and the values of
$(c_1,c_2$) are (1.0,1.0) for (a), (1.5,1.0) for (b), and (10000,10000) for (c).
(a) is the static case and (c) is a highly nonstatic one.
We have used $\epsilon=1$, $\hbar=1$, $k=1$, $V=1$, $\alpha_0=1$,
$t_0 = 0$, $\varphi = 0$, and $\theta=0$.
}
\end{figure}

We consider interference of two nonstatic coherent waves whose frequencies are different from each other.
Let us write the frequencies of the two waves as $\omega_{\rm I}$ and $\omega_{\rm II}$, respectively.
Then, the resultant electric field is expressed as
\be
E_{\rm tot} (x, t) = E_{\rm I} (x, t) + E_{\rm II} (x, t), \label{E215}
\ee
where
$E_{\rm I} (x, t) = E (x, t)|_{\omega = \omega_{\rm I}}$ and
$E_{\rm II} (x, t) = E (x, t)|_{\omega = \omega_{\rm II}}$, while $E (x, t)$ is
the one given in Eq. (\ref{20}).
The magnetic field can also be represented in a similar manner.
The time evolution of the field in Eq. (\ref{E215}), where $\omega_{\rm I}=1.0$ and $\omega_{\rm II}=1.5$,
is illustrated in Fig. 9.
We can see beating
from each panel of this figure, which was produced on account of the difference in frequency
between the two component fields.
Although the beating for the static wave is even through each position, it
becomes gradually uneven as the measure of nonstaticity increases.
The mean amplitude of the beating electric field is different depending on the position for nonstatic waves.
For instance, for the case of Fig. 9(b), the amplitude
is highest at $x=0.5$ and smallest at $x=2.0$ among the considered seven positions.
For a highly nonstatic wave given in Fig. 9(c),
the electric field is relatively very weak at $x=1.5$.
By the way, the phase difference between wave profiles whose positions are different
each other becomes negligible as the nonstaticity in the waves is being enhanced.
This is due to the fact that the nonstaticity is an effect associated with temporal variation of the wave,
but not related to its spatial variation.
However, the beating effect that happened due to such interference in the waves still remains
independently of the scale of nonstaticity.
\\
\\
{\bf 6. CONCLUSION AND OUTLOOK \vspace{0.2cm}} \\
We have investigated the effects of wave nonstaticity for light propagating in a static environment focusing
on the situation where the measure of nonstaticity is high.
We have shown that, if the measure of nonstaticity is extremely high,
the phase of the wave function evolves in a periodic step-like manner, i.e., it drops vertically
whenever the wave constitutes a node in the quadrature space.
The effects of such a step phase on the superposed quantum states
and wave interference, as well as on the evolution
of the associated electromagnetic waves, have been investigated.

The time evolution of the nonstatic electromagnetic wave is abnormal due to the above-mentioned peculiar
phase property in the wave functions.
The evolution profile of the phase factor in the electromagnetic wave over
time is a rectangular type in a highly nonstatic case.
However, the electric and magnetic components in the wave always evolve following the
sinusoidal form like those in an ordinary wave.
For the case that the wave nonstaticity is quite high, the electromagnetic waves are very much the same as the
standing wave where we can hardly know their propagation direction from their temporal behavior.

The novel abrupt drop of the phase of the wave function is originated from the enhancing of
the time-varying geometric phase as the nonstaticity of the wave augments.
We confirmed that the superposition state and the interference
of the electromagnetic waves in the nonstatic regime are very different from those in the ordinary wave due to
the effects of the uncommon temporal behavior of the phase.
In a superposition state, the detailed original superposition pattern in the evolution
of the associated probability density was gradually rubbed out as the nonstaticity grows and,
instead, the variation of the wave's amplitude that arose due to its nonstaticity was pronounced.

Effects of interference between two nonstatic electromagnetic waves which have different frequencies
were analyzed subsequently. In the extreme nonstatic limit in this analysis, the phase difference
in the resultant interfering electromagnetic wave between different positions is nearly not recognized
from the temporal wave profile by the dominance of the nonstaticity-induced characteristic in that profile:
this outcome is due to the fact that the wave nonstaticity treated here is a phenomenon associated entirely
with time variation of waves, while it is irrelevant to their spatial variation.

This work may provide a deeper understanding for the phase phenomena of nonstatic electromagnetic
waves which should be considered in wave manipulation in this context, especially for the phase modulation.
In order to utilize the nonstatic waves as a main resource in optical science and technology,
the demonstration of their rich properties associated with,
for example, the global phase coherence and its evolution that we have clarified may be necessary.
On one hand, the consequences of this research provide a tool for
characterizing how
the nonstatic fields affect light-matter interactions spatio-temporarily, which is in particular
important in the domain of ultrashort timescales \cite{wst1,wst2,wst3}.
\\
\appendix
\section{\bf Formula of Eigenfunctions}
The eigenfunctions in Eq. (\ref{1}) are represented in the form \cite{nwh}
\be
\langle q |\phi_n \rangle =
\left({\f{\zeta(t)}{\pi}}\right)^{1/4} \f{1}{\sqrt{2^n
n!}} H_n \left( \sqrt{\zeta(t)} q \right) \exp \left[
- \f{1}{2}\zeta'(t) q^2 \right], \label{qpn}
\ee
where $\zeta(t) = \epsilon\omega/[\hbar f(t)]$ and $\zeta'(t) = \zeta(t)-i\epsilon\dot{f}(t)/[2\hbar f(t)]$.
\\
\section{\bf Measure of Nonstaticity}
The measure of nonstaticity for nonstatic light waves in the Fock states are represented as \cite{nwh}
\be
D_{\rm F} = \f{\sqrt{(c_1+c_2)^2-4}}{2\sqrt{2}}.  \label{22}
\ee
The nonstaticity measure in the coherent state is also given by this relation \cite{ncs}.
\\
\section{\bf The Dynamical and the Geometric Phases}
The analytical formulae of the dynamical phase $\gamma_{D,n}(t)$ and the geometric
phase $\gamma_{G,n}(t)$ are given by \cite{gow}
\ba
\gamma_{D,n}(t) &=& -\f{1}{2}\left( n+\f{1}{2} \right)(c_1+c_2)\omega (t-t_0)+\gamma_{D,n}(t_0),
 \label{23} \\
\gamma_{G,n}(t) &=& \f{1}{2} \left( n+\f{1}{2} \right)
\{(c_1+c_2)\omega (t-t_0) -2 \omega T(t)\} +\gamma_{G,n}(t_0). \label{24}
\ea


\end{document}